\documentstyle[psfig,twocolumn,prl,aps,amssymb]{revtex}
\begin{document}
\wideabs{
\draft

\title{Charged Excitons of Composite Fermions in the Fractional
Quantum Hall Effect}
\author{Kwon Park}
\address{Department of Physics, Yale University, P.O. Box 208120,
New Haven, CT 06520-8120}
\date{\today}

\maketitle

\begin{abstract}
Charged excitons of composite fermions (CFs) are considered 
in the fractional quantum Hall effect. 
Energies of the charged CF excitons  
are computed at $\nu=1/3$
as a function of the size of exciton.
We show that
the charged CF exciton with size of roton  
is lower in energy than the unbound state of 
a neutral roton and a lone charge.
Therefore we propose that
the lowest-energy excitation of the fractional quantum Hall effect 
is in fact due to the charged excitons of composite
fermions composed of two CF-particles and one CF-hole. 
We believe that charged excitons of composite fermions shed new light on  
interpreting the resonant inelastic light scattering experiments.  
\end{abstract}

}


Two-dimensional electron systems (2DES)
exhibit spectacular phenomena such as the
fractional quantum Hall effect \cite{Tsui} (FQHE) when subjected to 
an intense, perpendicular magnetic field.
Excitations of the fractional quantum Hall effect have 
attracted considerable interest because of their key role in the 
existence of the fractional quantum Hall effect.
Now there has been much progress in understanding 
the excitations of FQHE in terms of the composite fermion (CF) theory 
\cite{Jain,Review1,Review2,Kamilla,Scarola,TwoRoton}.
Excitation gaps at a general momentum 
as well as the transport gap
have been described in terms of a single pair of CF-particle and CF-hole,
and shown to be 
in excellent agreement with numerous experiments and numerical studies.
Also, 
excitations with more complex structure than a simple, neutral exciton 
have proven
to be very important in the long-wavelength limit \cite{TwoRoton,Moonsoo}.
Originally, 
the question of whether the lowest-energy excitation at long wavelengths 
was described by a single exciton
or was instead a two-roton excitation was raised by 
Girvin, MacDonald, and Platzman \cite{GMP}, and later was considered
in numerical studies of finite systems \cite{He}. 
More recently, a conclusive evidence has been obtained 
from the composite fermion theory
in which the two-roton state was explcitly constructed and 
was shown to be lower in energy than 
a single CF exciton in the long wavelength limit \cite{TwoRoton}. 

\begin{figure}
\centerline{\psfig{figure=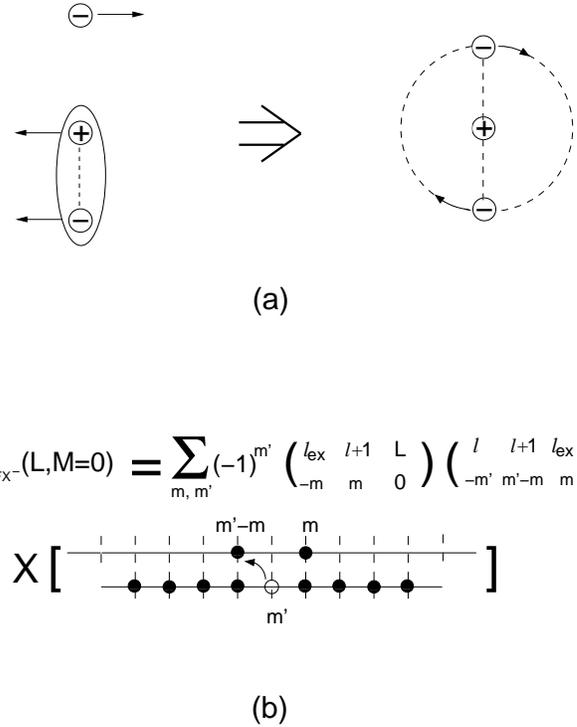,width=4.0in,angle=0}}
\caption{(a) Schematic diagram for the creation of   
a negatively charged exciton of composite fermions ($CFX^-$) with $L=0$.
(b) Schematic diagram for the wavefunction of $CFX^-$,   
with well defined $L$ and $M$ quantum numbers.
(We have chosen $M=0$ with no loss of generality, 
since the energy is independent of $M$.) 
The figure in square brackets shows schematically  
the Slater determinant obtained by promoting one  
composite fermion from the topmost fully occupied  CF Landau level 
to the next higher CF Landau level in single CF basis states indicated.  
(Note that the single CF basis states for this Slater determinant are 
the CF correlated basis functions which contains correlations 
from all other particles.) 
The topmost fully occupied LL corresponds to the angular momentum
$l$ shell in the spherical geometry; other Landau level shells are not 
shown for simplicity. 
The Wigner 3-j symbols are used in order to make a definite
angular-momentum eigenstate.  The relative signs of the
various terms in the sum follow from the antisymmetry requirement.
For a given $L$, there are in general multiple states 
to be constructed according to the angular-momentum addition rule 
between the angular momentum of single composite fermion ($l+1$) 
and that of neutral CF exciton ($l_{ex}$). 
\label{fig1}}
\end{figure}

However, there still exist possibilities 
for yet another excitation
partly because of the variational 
nature of the study. 
But more importantly, {\it when isolated quasi-particles 
or quasi-holes are already present in the system, 
it is possible that
an exciton takes advantage of them to form 
a bound state.}
Among the best candidates for the new excitation
is the charged exciton state of 
composite fermions which is composed of two CF-particles and 
one CF-hole (denoted as $CFX^-$), 
or one CF-particle and two CF-holes (denoted as $CFX^+$). 
Figure \ref{fig1}(a) shows a schematic diagram 
for the physical situation of 
interest in this article. 
A lone charge is combined with a nearby, neutral excitation
of opposite momentum to form a stationary, charged exciton 
of composite fermion. 
(For clear physical picture, consider that the lone charge 
is moving under the Landau gauge where each basis describes 
a plane wave along, say, $x$-direction.)

In fact, there has been considerable interest in the charged excitons
of electrons in two spatial dimension becuase of the increased binding 
energy due to the reduced dimensionality \cite{Lampert,Shields1,Finkelstein}. 
Even the charged excitons in strong magnetic fields have been studied 
by a number of authors, all of whose works, however, concentrated on
the excitons between electrons in the conduction band and holes in
the valence band \cite{MacDonald,Wojs1,Wojs2,BarJoseph}.  
On the contrary, 
the charged excitons studied 
in this article are formed by quasi-particles and quasi-holes 
of composite fermions which are 
in the same layer, band and electronic Landau level. 
We wish to show that these charged excitons of composite fermions are 
resposible for the lowest excitations of FQHE.

A composite fermion (CF) is the bound state of 
an electron and an even number of
magnetic flux quanta (a flux quantum is defined as $\phi_0=hc/e$), formed when 
electrons confined to two dimensions are exposed to a strong magnetic field
\cite{Jain,Review1,Review2}.  According to this theory, 
the interacting electrons at the Landau level (LL) filling factor 
$\nu=n/(2pn\pm 1)$, $n$ and $p$ being integers, transform into weakly
interacting composite fermions at an effective filling $\nu^*=n$; 
the ground state corresponds to $n$ filled CF Landau levels (CF-LLs) 
and it is natural to expect 
that the neutral excitations correspond to a particle-hole pair of 
composite fermions,  called the (neutral) CF exciton (denoted as $CFX^0$).
At the minimum in its dispersion, the CF exciton 
is called the roton, borrowing 
the terminology from the $^4$He literature \cite{GMP}.
Addition to the remarkable success of the composite fermion theory for the
FQHE states at the filling factors $\nu=n/(2pn\pm 1)$, it is 
equally amazing that the composite fermion theory 
continues to provide 
excellent states even away from those special filling factors. 
This broad accuracy enables us 
to utilize the composite fermion theory 
in order to study the systems with 
additional particles or holes on top 
of the filled CF-Landau level, the exicted states of which
are described by the charged excitons of composite fermions.

We will use the spherical geometry \cite{Monopole} below, which considers $N$
electrons on the surface of a sphere 
in the presence of a radial magnetic field emanating
from a magnetic monopole of strength $Q$, 
which corresponds to a total flux of $2Q\phi_0$
through the surface of the sphere.
The wave function for the CF state at $Q$, denoted by $\Psi_{2Q}$, 
is constructed by mapping to the wave function of
the corresponding electron states at $q$, denoted by $\Phi_{2q}$:
\begin{equation}
\Psi_{2Q}={\cal P}_{LLL} \Phi_{N-1}^{2p}\Phi_{2q}
\label{cfmapping}
\end{equation}
Here $\Phi_{N-1}=\prod_{j<k}(u_j v_k - u_k v_j)$ 
is the wavefunction of the fully occupied
lowest Landau level with monopole strength equal to $(N-1)/2$, 
where $u_j \equiv \cos(\theta_j /2)
\exp(-i\phi_j /2)$ and  $v_j \equiv \sin(\theta_j /2) \exp(i\phi_j /2)$.
${\cal P}_{LLL}$ denotes projection of the wave function 
into the lowest Landau level (LLL).  
The monopole strengths for $\Phi_{2q}$ and $\Psi_{2Q}$, $q$ and $Q$, 
respectively, are related by $Q=q+p(N-1)$. 
It is crucial to note that
formally the mapping between $\Psi_{2Q}$ and $\Phi_{2q}$ 
can be defined regardless of whether or not 
they are incompressible. The accuracy of $\Psi_{2Q}$ against
the exact state, however,
is dependent on whether $\Phi_{2q}$ is robust when subjected to
perturbations such as residual interactions between composite fermions.
In particular, for the ground state and the single exciton state of 
the system with completely filled CF-Landau levels, 
the wave functions $\Phi_{2q}$ are completely determined by
symmetry (i.e., by fixing the total orbital angular momentum $L$, 
which is preserved in 
going from $\Phi_{2q}$ to $\Psi_{2Q}$ according to Eq.(\ref{cfmapping})), 
giving parameter-free
wave functions $\Psi_{2Q}$ for the ground and excited states of 
interacting electrons.
These have been found to be extremely accurate in tests against exact
diagonalization results available for small systems \cite{Jain,JK}, 
establishing the essential validity of the CF exciton description of the
neutral mode of the FQHE.

Figure~\ref{fig1}(b) shows a schematic diagram for the construction of
the negatively charged CF exciton. Unlike the previous case of
completely filled CF Landau levels, there are in general multiple states 
to be constructed for a given value of angular momentum $L$
according to the angular-momentum addition rule 
and the Fermi statistics. 
In fact, the distinction between
the lone charge and the constituent charge from the neutral exciton 
is arbitrary because thier contribution to the final charged
exciton state must be identical. 
But whenever there is a unique state for $\Phi_{2q}$, 
determined only by symmetry, the chioce of intermediate route for 
constructing the wavefunction does not affect 
the final state of the charged CF exciton. More importantly, 
its uniqueness suggests the robustness of $\Phi_{2q}$, and therefore
$\Psi_{2Q}$. 
However, when 
there are multiple states for $\Phi_{2q}$ in the same angular
momentum channel, it is necessary to diagonalize the Hamiltonian 
in the restricted Hilbert space, 
the basis functions of which are provided by the 
composite fermion theory. In this case, the final composite fermion 
wavefunction is more susceptible 
to the form of interactions between particles.

\begin{figure}
\centerline{\psfig{figure=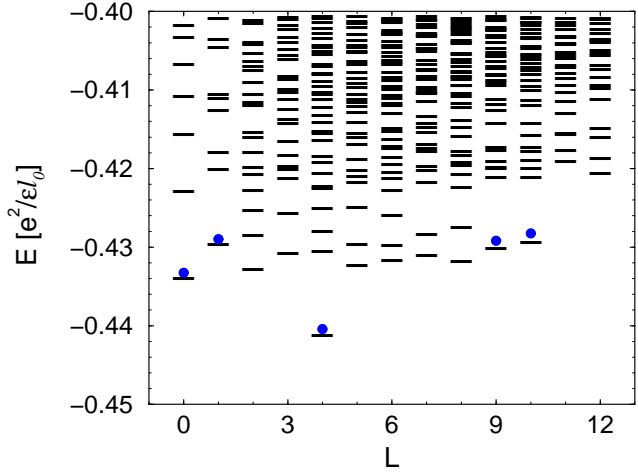,width=4.0in,angle=-90}}
\caption{Comparison between the energies 
of the ground state (isolated charge) at $L=4$ and
the lowest lying excitations (charged CF excitons) 
at $L=0$, $1$, $9$, and $10$ 
in the composite fermion theory, and those of the exact diagonalization study
for the system of the number of electrons $N=8$ 
and the monopole strength $Q=10$. 
Energies from the composite fermion theory 
are denoted as solid circles 
while the exact energies are indicated
by short horizontal lines. 
Note that the Coulomb interaction 
is taken for the interaction between electrons. 
Also, note that 
the statistical uncertainty from the Monte Carlo simulation is 
smaller than the size of symbols.
\label{fig2}}
\end{figure}

Therefore it is very convenient that 
the $L=0$ state of charged CF exciton  
is uniquely determined, which, in turn, guarantees      
that correlations between CF-particles and CF-holes are automatically
taken into account because of the exactly same reason as for the neutral 
excitons, $CFX^0$. Figure~\ref{fig2} shows the comparison between
the composite fermion states and the exact states for the system of
number of elercrons $N=8$ and the monopole strength $Q=10$, 
where the ground state describes a lone, negatively charged
composite fermion at $\nu=1/3$. 
Addition to the $L=0$ case mentioned above, 
a unique chioce is obtained for the charged CF exciton state at 
$L=1$, $9$, and $10$.
(Note that the Fermi statistics is important to determine 
the number of states in each angular-momentum channel since two
particles are in the same CF Landau level.)
Energies per particle for the charged CF exciton state at 
$L=0$, $1$, $9$, and $10$
are respectively
$-0.433257(63)$, $-0.428987(60)$, $-0.429184(40)$, and $-0.428253(57)$ 
in units of $e^2 /\epsilon l_0$, 
which agree with the exact energies,
$-0.433592$, $-0.429188$, $-0.429676$, and $-0.428982$,     
within less than $0.2 \%$. 
Also, the energy of the ground state at $L=4$
in the composite fermion theory, $-0.440415(61)$, 
is in excellent agreement with 
the exact ground state energy, $-0.440764$.

Since it is established that the composite fermion theory
provides accurate states for the charged exciton state, 
we next turn to the energetics. 
But, before we go into the detail, several comments are in order.
First, though the formalism of constructing the charged CF exciton state
can be applied to general filling factors, we restrict our attention to 
the charged CF excitons at $\nu=1/3$.
Second, it is assumed 
that the Zeeman splitting energy is large enough to suppress the 
spin-flip excitations. 
Third, we only consider the negatively charged CF exciton
with $L=0$ which is, in other words, a stationary exciton. 
Note that in general a non-zero momentum-transfer is required 
to form a charged exciton since the ground state
of a lone charge is not a uniform state.
And finally, regarding the technical aspect 
of Monte Carlo simulation, the efficient
determinant-updating technique in Ref.~\cite{TwoRoton} has been used 
because of the large number of Slater determinants required in constructing
the charged CF exciton state.

Now, the energy cost of creating a stationary, charged CF exciton, 
is defined as follows:
\begin{eqnarray}
\Delta E_{CFX^-}(L_{ex}) = E_{CFX^-}(L=0) - E_{lone\;\;CF}(L=L_{gr})
\end{eqnarray}
where $L_{gr}$ is the angular momentum of the ground state
containing the lone composite fermion. 
For example, $L_{gr}=4$ for the system of $N=8$ and $Q=10$. 
And $L_{ex}$ is the angular momentum of the constituent single
exciton, which is identical to $L_{gr}$ 
since we are only interested in the stationary charged exciton, 
i.e. $L=0$. Note that, as usual, 
the size of constituent single exciton, 
and therefore the charged exciton, is proportional to 
$k_{ex} l^2_0$ 
where $k_{ex}=L_{ex}/R$ and $R=\sqrt{Q} l_0$.
However, the above relationship between the angular momentum in
the spherical geometry and the linear momentum in planar geometry
is valid only for the neutral exciton.
Also note that it is not possible to increase the number of electrons 
in order to reach the thermodynamic limit 
for a fixed value of $k_{ex} l_0$ because of the nature of 
the spherical geometry. So $k_{ex} l_0$ is also changed
upon increasing the system size. Of course, 
it is possible in priciple to construct a charged CF exciton
for an arbitrary number of electrons by using the diagonalization
within the restricted CF Hilbert space.
However, this procedure is very hard to be carried out in practice.

The energy cost of excitation far away from the lone charge,
or $CFX^{0}+CF$, 
is identical to that of neutral excitation from the filled CF shell: 
\begin{eqnarray}
\Delta E_{CFX^{0}+CF}(L_{ex}) &=& E_{CFX^0}(L=L_{ex})
\nonumber \\ 
&-& E_{filled\;\;shell}(L=0)
\end{eqnarray}
Therefore it is natural to define the binding energy of the charged
CF exciton,
$\Delta_{CFX^-}$, as follows:
\begin{eqnarray}
\Delta_{CFX^{-}}(L_{ex}) = \Delta E_{CFX^{0}+CF}(L_{ex})
- \Delta E_{CFX^{-}}(L_{ex}) 
\end{eqnarray}

\begin{figure}
\centerline{\psfig{figure=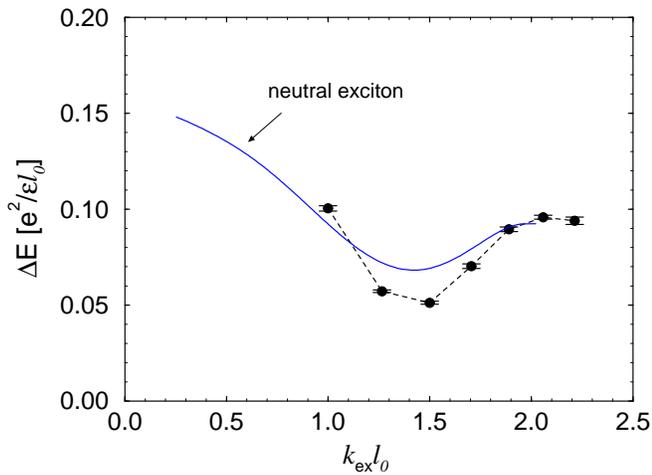,width=4.0in,angle=-90}}
\caption{Energy of the charged CF excitons  
as a function of the momentum ($k_{ex} l_0$) 
of the constituent single exciton.
Note that $k_{ex} l^2_0$ is proportional to the distance
between the CF-particle and the CF-hole of the constituent single exciton, 
and therefore it measures the 
size of the charged CF exciton. 
As shown from the comparison with the dispersion curve of neutral exciton,
the energy of the charged CF exciton is quite lower than that of
single exciton around $k_{ex} l_0= 1.5$, i.e. 
{\it by combining a lone charge and a single roton}.
\label{fig3}}
\end{figure}

Figure~\ref{fig3} shows the Coulomb energies of 
the charged CF exciton as a function of $k_{ex} l_0$, compared with
those of the neutral excitons. 
As shown in Fig.\ref{fig3}, the charged CF exciton is quite
lower in energy than the single exciton state 
around $k_{ex} l_0 = 1.5$. 
In other words, the charged CF exciton has 
the lowest energy when created by {\it combining a lone charge and a roton}. 
The binding energy of the charged CF exciton is estimated to be
roughly $0.02 e^2/\epsilon l_0$. 
As expected, for large $k_{ex} l_0$, 
the energy of charged CF exciton is equal to
that of single exciton since 
the charged exciton with large $k_{ex} l_0$
means the far-separated, independent collection of 
quasi-particles and quasi-holes.

Now let us turn to the relevance of this work to the resonant
inelastic light scattering experiments. 
A conventional interpretation of the excitations measured in those
experiments is that they are neutral exictons. However, since there are
always some lone charges in the real system of experiments, 
it is plausible
that the light-scattering gives rise to the charged CF excitons. 
As shown in the above, the charged CF exciton can have 
a lower energy than the single, neutral roton. Moreover,
by using the analogy to 
the charged excitons of electrons \cite{Shields},
the charged CF exciton can decrease its energy further 
relative to the neutral excitons because it can be bound to 
a donor. The distance between 2DES and the donor layer
is not very far when measured in units of magnetic length: 
typically less than 10 times magnetic length. Therefore 
the interaction between the charged exciton and a donor 
can lower the energy significantly, 
bringing theory and experiments closer. 

However, a more interesting consequence is that the energy reduction 
due to the interaction with donors is constant in 
laboratory units such as $MeV$, but is increasing in units of
$e^2 /\epsilon l_0$ as the electron density becomes smaller.
This may explain the recent, curious experimental observation\cite{Moonsoo} 
that the energies of the roton as well as the long-wavelength excitation
is almost constant as a function of electron density 
whereas the CF theory combined with 
the local density approximation \cite{ParkLDA,Scarola} 
predicts the increase of energies as the electron density decreases 
because of the smaller role of the finite thickness effect. 
The energy reduction due to the interaction with donors is bigger 
for a smaller density, and therefore bring down the excitation energies
more. Finally, it is speculated that the long-wavelength excitation
can be composed of two charged CF excitons at their minimum energy,
which is analogous to the two-roton state.

In conclusion, 
we have shown that the charged CF exciton 
composed of two CF-particles and one CF-hole is lower in energy
than the sigle roton, when created by combining a lone charge and a roton,
and its binding energy is estimated to be 
around 0.02 $e^2/\epsilon l_0$. 
Therefore it is proposed that
the lowest excitation of the FQHE state at $\nu=1/3$,
is in fact due to the charged exciton of composite fermions.
Study of the relevance of the charged CF exciton 
at other filling factors is in progress.

This work was supported in part by the National Science
Foundation under Grant No. DMR-9986806 and DMR-9800626.
The author is very grateful to Jainendra Jain for his 
insightful discussion and constant encouragement, and Nick Read 
for his helpful comments. 
Also, the author is deeply indebted to R. Shankar and Subir Sachdev 
for their support in Yale.

\end{document}